\documentclass[12pt]{article}
\usepackage{amsfonts,a4}
\usepackage{pictex}
\usepackage{graphics}
\usepackage{amsmath}
\usepackage{amssymb}
\newcommand{\nc}{\newcommand}

\newcommand{\diam}{{\rm{diam}}\,}
\renewcommand{\Im}{\mathrm{Im}\,}
\nc{\di}{\displaystyle}
\nc{\nn}{\nonumber}
\nc{\nek}{\nonumber\\[1ex]}
\nc{\st}{\scriptstyle}
\nc{\tst}{\textstyle}
\nc{\nb}{\normalsize\bf}
\nc{\ns}{\normalsize}
\nc{\seq}{\subseteq}
\nc{\AH}{\mathcal{A}}
\nc{\BE}{\mathcal{B}}
\nc{\CE}{\mathcal{C}}
\nc{\D}{\mathcal{D}}
\nc{\E}{\mathcal{E}}
\nc{\EF}{\mathcal{F}}
\nc{\GE}{\mathcal{G}}
\nc{\HA}{\mathcal{H}}
\nc{\J}{\mathcal{J}}
\nc{\KA}{\mathcal{K}}
\nc{\EL}{\mathcal{L}}
\nc{\PE}{\mathcal{P}}
\nc{\ER}{\mathcal{R}}
\nc{\ES}{\mathcal{S}}
\nc{\TE}{\mathcal{T}}
\nc{\EM}{\mathcal{M}}
\nc{\EN}{\mathcal{N}}
\nc{\OH}{\mathcal{O}}
\nc{\U}{\mathcal{U}}
\nc{\WE}{\mathcal{W}}
\nc{\EX}{\mathcal{X}}
\nc{\Y}{\mathcal{Y}}
\nc{\ZE}{\mathcal{Z}}
\nc{\ma}[1]{\mbox{$\,{#1}\,$}}
\nc{\ek}{\protect\\[1ex]}
\nc{\zx}{\protect\\[2ex]}

\newcommand{\R}{{\mathbb R}}

\nc{\IR}{\mbox{\bf R}}
\nc{\IN}{\mbox{\bf N}}
\nc{\ZZ}{\mbox{\bf Z}}
\nc{\la}{\lambda}
\nc{\La}{\Lambda}
\nc{\da}{\delta}
\nc{\Da}{\Delta}
\nc{\ta}{\theta}
\nc{\Ta}{\Theta}
\nc{\na}{\nabla}
\nc{\ue}{\infty}
\nc{\vp}{\varphi}
\nc{\vta}{\vartheta}
\nc{\Gm}{\Gamma}
\nc{\gm}{\gamma}
\nc{\ka}{\kappa}
\nc{\si}{\sigma}
\nc{\Si}{\Sigma}
\nc{\al}{\alpha}
\nc{\be}{\beta}
\nc{\om}{\omega}
\nc{\Om}{\Omega}
\nc{\pa}{\partial}
\nc{\ti}{\times}
\nc{\n}{|}
\nc{\rub}{\,\rule[-2.7pt]{.02in}{4mm}\,}
\nc{\ab}{\|}
\nc{\s}{\tilde}
\nc{\ve}{\varepsilon}
\nc{\fa}{\forall}
\nc{\ov}{\overline}
\nc{\un}{\underline}
\nc{\Llr}{\Longleftrightarrow}
\nc{\llr}{\longleftrightarrow}
\nc{\ra}{\rightarrow}
\nc{\lra}{\longrightarrow}
\nc{\rh}{\rightharpoonup}
\nc{\Ra}{\Rightarrow}
\nc{\ran}{\rangle}
\nc{\lan}{\langle}
\nc{\bs}{\backslash}
\nc{\ko}{\,,\,}
\nc{\eq}[1]{\mbox{\rm {(\ref{E#1})}}}
\nc{\qed}{\mbox{ }\nolinebreak\hfill \rule{2mm}{2mm}}
\nc{\ha}{\frac{1}{2}}
\nc{\lk}{\left[}
\nc{\rk}{\right]}
\nc{\lb}{\left\{}
\nc{\rb}{\right\}}
\nc{\rr}{\right)}
\nc{\lr}{\left(}
\nc{\f}{\big(}
\nc{\g}{\big)}
\nc{\Ba}{\Big(}
\nc{\Bz}{\Big)}
\nc{\Bka}{\Big[}
\nc{\Bkz}{\Big]}
\nc{\bka}{\big[}
\nc{\bkz}{\big]}
\nc{\Blb}{\Big\{}
\nc{\Brb}{\Big\}}
\nc{\blb}{\big\{}
\nc{\brb}{\big\}}
\nc{\pn}{\par\noindent}
\nc{\emp}{\emptyset}
\nc{\Ri}{\Rightarrow}
\nc{\rain}[1]{\raisebox{-3pt}{${\tst |}_{#1}$}}
\nc{\hph}{\hphantom}
\nc{\vph}{\vphantom}
\nc{\vpn}{\vspace{2ex}\par\noindent}
\nc{\vpar}{\vspace{2ex}\par}
\nc{\mathe}[1]{\mbox{${\di {#1}}$}}
\nc{\sD}{{\tilde{D}}}
\newtheorem{lem}{Lemma}
\newtheorem{theo}{Theorem}

\newtheorem{rem}{Remark}
\newtheorem{exa}{Example}

\renewcommand{\theequation}{\thesection.\arabic{equation}}
\numberwithin{equation}{section}
\begin{document}
\title{{\Large {\bf Many-body wave scattering\\ by small bodies and applications
}}}
\author{A. G. Ramm \\ {\ns (Mathematics Department, Kansas St. 
University,}\\ {\ns Manhattan, KS 66506, USA} \\ {\ns and TU Darmstadt, 
Germany)}\\ {\small ramm@math.ksu.edu}} \date{}
\maketitle
\pn{\small {\em PACS:} 03.04.Kf, 43.20.tg, 62.30.dt
}
\begin{abstract}
A rigorous reduction of the many-body wave scattering problem to solving
a linear algebraic system is given bypassing solving the usual system of
integral equation. The limiting case of infinitely many small particles
embedded into a medium is considered and the limiting equation for the
field in the medium is derived. The impedance boundary conditions are
imposed on the boundaries of small bodies. The case of  Neumann boundary
conditions (acoustically hard particles) is also considered.
Applications to creating materials with a desired refraction coefficient
are given. It is proved that by embedding suitable number of small 
particles per unit volume of the original material with suitable boundary 
impedances one can create a new material with any desired refraction 
coefficient. The governing equation is a scalar Helmholtz equation,
which one obtains by Fourier transforming the wave equation. 

\end{abstract}
\section{Introduction}\label{SI}
This paper is a continuation of \cite{R518}
and uses some of the results from \cite{R476}, \cite{R450}, \cite{R506},
\cite{R524}, \cite{R525}. Applications of our theory to creating
materials with desired refraction coefficient, including negative
refraction, are discussed in \cite{R513}, \cite{R515}, \cite{R516},
\cite{R527}. Wave scattering by small bodies is a classical branch of
science: it was originated by Rayleigh in 1871. In \cite{L} one finds
a discussion of wave scattering by a small particle.
In \cite{DK} there is a review of the low frequency scattering theory and
formulas for scattering by small balls and ellipsoids are given. In 
\cite{R476} the
theory is developed for small bodies of arbitrary shapes. In \cite{R518}
the many-body scattering problem was reduced to solving linear algebraic
systems bypassing the usual study of a system of integral equations. In
this paper we apply the approach proposed in \cite{R518} and study the
limiting behavior of the scattering solution when the number of small
bodies tends to infinity in such a way that the characteristic size $a$ of
the small particles is related to their number $M$ so that 
$M=O(\frac 1 a)$ in Theorem 2, and $M=O(\frac 1 {a^3})$ in Theorem 3. 
Sufficient conditions for convergence of the
scattering solution in this limiting process are given. We prove that
these conditions are, in some sense, also necessary for convergence. The 
limit of the
scattering solution is a function, which satisfies some differential
or integral-differential equations. These equations describe the
behavior of the wave field in the new medium, obtained in the limit.

There is a large literature on the calculation of the effective
dielectric permittivity and magnetic permeability of the 
composite materials (Maxwell-Garnett and Bruggeman recipes and their 
numerous versions, and newer theories \cite{Sh}, \cite{Mi}). 
In the literature mostly a randomly uniform distribution of the inclusions 
is assumed and the resulting homogenized medium is described by 
effective constant dielectric permittivity and magnetic permeability,
which can be tensors. In this work the propagation and scattering of 
scalar waves are discussed, and the "homogenized" medium is described 
not by a constant refraction coefficient, but by a refraction coefficient 
which is a function of spatial variables.   

Let us formulate the problem. Consider first a bounded domain $D\subset
\R^3$ filled with a material with a known refraction coefficient
$n_0(x)$. The governing equation is:
\begin{equation}
\label{E1.1}
L_0 u_0 := \f\na^2 + k^2 n_0(x)\g u_0 = 0\quad \mbox{in }\R^3.
\end{equation}
We assume that $n_0(x) = 1$ in $D' = \R^3\bs D$, $k = {\rm const}>0$, and
$n_0 = \max_{x\in D}\n n_0(x)\n < \ue$. The operator $L_0$ can be written
as a Schr\"odinger operator:
\begin{equation}
\label{E1.2}
L_0 = \na^2 + k^2 - q_0(x),\quad q_0(x):= k^2[1-n_0(x)],
\end{equation}
and $q_0 = 0$ in $D'$. One has 
$$n_0(x) = 1-k^{-2}q_0(x),$$ so there is a
one-to-one correspondence between $n_0(x)$ and $q_0(x)$. If $n_0(x)$ is
known, then one knows the scattering solution:
\begin{align}
\nn
L_0 u_0& = 0\quad \mbox{in }\R^3,\\
 u_0(x) & = e^{ik\al\cdot x} +
A_0(\be,\al)\, \frac{e^{ikr}}{r} + o\f \frac{1}{r}\g,\quad r = \n x\n
\ra \ue, \be := \frac{x}{r}.\label{E1.3}
\end{align}
The coefficient $A_0(\be,\al)$ is called the scattering amplitude, the
unit vector $\al\in S^2$ is given, $\al$ is the direction of the
incident plane wave $e^{ik\al\cdot x}$, $S^2$ is the unit sphere in
$\R^3$, $\be\in S^2$ is the direction of the scattered wave, $k > 0$ is
a wave number, which we assume fixed throughout the paper. By this
reason we do not show the $k$-dependence of $A$ and $u_0$.

Let $G(x,y)$ be the resolvent kernel of $L_0$ satisfying the radiation
condition (or the limiting absorption principle):
\begin{equation}
\label{E1.4}
L_0G(x,y) = -\da(x-y)\quad \mbox{in }\R^3.
\end{equation}
This function $G(x,y)$ is known because $q_0(x)$ is known.

Consider now the scattering problem for many small bodies $D_m$ embedded
in $D$, $1 \leq m \leq M$:
\begin{eqnarray}
\label{E1.5}
 L_0 u_M &=& 0 \quad \mbox{in } \R^3 \bs \bigcup^M_{m =1} D_m,\\
 u_M &=&  u_0 + A_M(\be,\al)\, \frac{e^{ikr}}{r} + o\f
\frac{1}{r}\g,\quad r = \n x\n \ra \ue,\quad \frac{x}{r} = \be,
\hspace{2cm}\label{E1.6}\\
 \frac{\pa u_M}{\pa N} &=& \zeta_m u_M\quad \mbox{on }S_m:= \pa
D_m,\quad 1\leq m\leq M,\label{E1.7}
\end{eqnarray}
where $u_0$ is the solution ot the scattering problem \eq{1.3}. Here $N$
is the normal to $S_m$ pointing out of $D_m$, $\zeta_m$ is a complex
number, the boundary impedace, ${\rm Im}\, \zeta_m\leq 0$, $S_m$ is  
uniformly  $C^{1,\la}$ with respect to
$m$, $1 \leq m \leq M$. By $C^{1,\la}$ surface we mean the surface
with local equation $x_3 = f(x_1,x_2)$, where
$f\in C^{1,\la}$, $\la > 0$. We assume throughout this paper that
\begin{equation}
\label{E1.8}
n_0 ka \ll 1,\quad d \gg a,
\end{equation}
where
\begin{equation}
\label{E1.9}
a = \ha\, \max_m\, {\rm diam} D_m,\quad d = \min_{m\neq j}\, 
{\rm dist}(D_m,D_j).
\end{equation}
By $V_m := \n D_m\n$ the volume of $D_m$ is denoted, and by $\n S_m\n$
the surface area of $S_m$ is denoted.

One can prove (see Section \ref{S3}) that problem \eq{1.5} -- \eq{1.7}
has at most one solution if ${\rm Im}\, \zeta_m \leq 0$, $1 \leq m \leq
M$, and ${\rm Im}\, q_0(x)\leq 0$. 

We look for the solution to problem
\eq{1.5} -- \eq{1.7} of the form
\begin{equation}
\label{E1.10}
u_M(x) = u_0(x) + \sum^M_{m=1} \int_{S_m} G(x,s)\si_m(s)ds,
\end{equation}
where $\si_m$ should be found from the boundary conditions \eq{1.7}. For
any $\si_m$ the function \eq{1.10} solves equation \eq{1.5} and
satisfies condition \eq{1.6}:
\begin{equation}
\label{E1.11}
A_M(\be,\al) = \frac{1}{4\pi} \sum^M_{m=1} \int_{S_m} u_0(s,{-}\be)\si_m(s)ds.
\end{equation}
Formula \eq{1.11} follows from \eq{1.6}, \eq{1.10} and the Ramm's
lemma (\cite{R470}, formulas (5.1.31), (5.1.36)):
\begin{equation}
\label{E1.12}
G(x,y) = \frac{e^{ik\n x\n}}{4\pi \n x\n}\, u_0(y,\alpha) + o \Ba
\frac{1}{\n x\n}\Bz,\quad \n x\n \to \ue,\: \frac{x}{\n x\n} = {-}\alpha,
\end{equation}
where $u_0(x,\alpha)$ is the scattering solution.
A similar formula was proved earlier in \cite{R190}, p. 46, for the
resolvent kernel of the Laplacian in the exterior of a bounded obstacle,
(and even earlier, in \cite{R3}, for some unbounded obstacles). The
scattering amplitude for problem \eq{1.5} -- \eq{1.7} is
\begin{equation}
\label{E1.13}
A(\be,\al) = A_0(\be,\al) + A_M(\be,\al),
\end{equation}
where $A_0$ is defined in \eq{1.3} and $A_m$ is defined in \eq{1.6}. If
$ka$ is sufficiently small, then $k^2$ is not a
Dirichlet eigenvalue of the operator $\nabla^2-q_0(x)$ in 
$D_m$, $1\leq m \leq M$. If
\begin{equation}
\label{E1.14}
{\rm Im}\, \zeta_m\leq 0,\quad 1 \leq m \leq M;\quad \Im q_0(x)\leq 0,
\end{equation}
then the unique solution to problem \eq{1.5} -- \eq{1.7} can be found in
the form \eq{1.10}.
\begin{theo}\label{T1}
Assume \eq{1.8} and \eq{1.14}. Then problem \eq{1.5} --
\eq{1.7} has a solution of the form \eq{1.10} and this is
the unique solution of the problem \eq{1.5} --\eq{1.7}.
\end{theo}
{\bf Proof of Theorem \ref{T1}} is given in Section \ref{S3}. In \cite{MK}
there is a detailed study of boundary value problems in domains of the
type $D':=\R^3\bs \bigcup^M_{m=1} D_m$. In \cite{MK} the case of Dirichlet
boundary condition on $S_m$ was studied, the case of Neumann boundary
condition was mentioned as an open problem, and the case of impedance
boundary condition was not studied. 

Let
\begin{equation}
\label{E1.15}
g(x,y):= \frac{e^{ik\n x-y\n}}{4\pi \n x-y\n},\quad g_0(x,y):=
\frac{1}{4\pi \n x-y\n}\,.
\end{equation}
Note that
\begin{equation}
\label{E1.16}
G(x,y) = g(x,y)-\int_D g(x,z)q_0(z)G(z,y)dz.
\end{equation}
We need two lemmas.
\begin{lem}\label{L1}
If
\begin{equation}
\label{E1.17}
\n t-x\n \leq a,\quad \n x-y\n \geq d \gg a,
\end{equation}
then
\begin{equation}
\label{E1.18}
\n g(t,y) - g(x,y)\n \leq c \Ba \frac{a}{d^2} + \frac{ka}{d}\Bz,
\end{equation}
where $c > 0$ stands for various positive constants independent of $a$
and $d$.
\end{lem}
\begin{lem}\label{L2}
If \eq{1.17} holds, then
\begin{equation}
\label{E1.19}
\n G(t,y)-G(x,y)\n \leq c \Ba \frac{a}{d^2} + \frac{ka}{d}\Bz.
\end{equation}
\end{lem}
These lemmas are proved in Section \ref{S3}.

Let us formulate our results under simplifying but physically reasonable
assumptions.
\begin{theo}\label{T2}
Assume that
\begin{equation}
\label{E1.20}
\lim_{\substack{a\ra 0\\ x_m\in D_m}}\, \frac{\zeta_m J_m}{4\pi \n
S_m\n} = h(x_m), \quad \mbox{where } J_m:= \int_{S_m}\int_{S_m}
\frac{dsdt}{\n s-t\n}\,,
\end{equation}
and for any subdomain $\s{D}\subset D$ the following relation holds
\begin{equation}
\label{E1.21}
\sum_{D_m\subset \s{D} } 1 = \frac{1}{a} \int_{\s{D}} N(x)dx.
\end{equation}
Assume that $\n S_m\n = c_1 a^2$,  and $J_m = c_2 a^3$, where $c_1,c_2 >
0$ are constants independent of $m$.

Finally we assume that $M = O(a^{-1})$ and $d = O(a^{1/3})$ as $a\ra 0$.
Under these assumptions there exists the limit:
\begin{equation}
\label{E1.22}
\lim_{M\ra \ue} u_M(x) = u(x)=u(x,\alpha).
\end{equation}
This $u(x)$ solves the equations:
$$u(x) =  u_0(x) - \int_D G(x,y)\, p(y) \, u(y)dy, $$
and
\begin{equation}
\label{E1.23}
L u := [\na^2 + k^2 - q(x)]u = 0\quad \mbox{in } \R^3,
\end{equation}
where the potential $q$ is of the form:
\begin{equation}
\label{E1.24}
q(x) = q_0(x) + p(x),\quad p(x) = \frac{4\pi c_1^2 \, N(x) h(x)}{c_2 [1
+ h(x)]}\,,
\end{equation}
and $u$ satisfies the radiation condition:
\begin{equation}
\label{E1.25}
u = e^{ik\al\cdot x} + A(\be,\al)\, \frac{e^{ikr}}{r} + o\Ba
\frac{1}{r}\Bz,\quad r = \n x\n\ra \ue,
\end{equation}
where 
\begin{equation}
\label{E1.26a}
A(\be,\al) = A_0(\be,\al) - \frac{1}{4\pi} \int_D u_0(y,{-}\be) 
p(y)u(y,\alpha)dy,
\end{equation}
and $u_0(y,-\beta)$ is the scattering solution defined in \eq{1.3}.
\end{theo}
\begin{theo}\label{T3}
Assume that $\zeta_m = 0$, $1\leq m \leq M$, and the following limits
exist:
\begin{eqnarray}
\label{E1.26b} \lim_{a\ra 0} \sum_{D_m\subset \s{D}} V_m 
\be^{(m)}_{pj}& =&
\int_{\s{D}} \be_{pj}(y)\nu (y)dy,
\\ \label{E1.27}
\lim_{a\ra 0} \sum_{D_m\subset \s{D}}V_m &=& \int_{\s{D}} \nu(y)dy,
\end{eqnarray}
where $\nu(y)\geq 0$ and $\be_{pj}(y)$ are continuous functions in
$D$, and $\be^{(m)}_{pj}$ is the magnetic polarizability tensor of the 
body $D_m$, defined in \eq{2.38}-\eq{2.39}, see below.

Then the function $u_M(x)$, defined in \eq{1.10}, tends to the
limit:
\begin{equation}
\label{E1.28}
\lim_{M\ra \ue}u_M(x) = \U(x)=\U(x, \alpha),
\end{equation}
and $\U(x)$ solves the equation:
\begin{equation}
\label{E1.29}
\U(x) = u_0(x) + \int_D G(x,y)\Da\, \U(y)\nu (y)dy - 
\sum^3_{p,j=1}\int_D
\frac{\pa G(x,y)}{\pa y_p}\; \frac{\pa \U(y)}{\pa y_j}\, \be_{pj}(y)\nu(y)dy.
\end{equation}
\end{theo}
If all the small particles are balls of radius $a>0$, then 
$$V_m=\frac {4\pi a^3}{3},\quad |S_m|=4\pi a^2, \quad J_m=16\pi^2 a^3,
\quad \int_{S_m}\frac {dt}{4\pi|s-t|}=a, \,\, s\in S_m.$$
In this case 
$$\int_{S_m}\frac {dt}{|s-t|}=\frac 1 
{|S_m|}\int_{S_m}\int_{S_m}\frac {dtds}{4\pi|s-t|},$$
that is, the mean value of the integral $\int_{S_m}\frac {dt}{|s-t|}$
on the surface $S_m$ equals to this integral. If $S_m$ is not a sphere,
this mean value is an approximate value of the above integral. 

Note that under the assumptions of Theorem \ref{T2} one has $M =
O(a^{-1})$, while under the assumptions of Theorem \ref{T3} one has $M =
O(a^{-3})$ (see formula \eq{2.50} below). Therefore, one needs many more
particles to deal with the Neumann boundary condition, that is, with
acoustically hard particles, than with the impedance boundary condition
with large boundary impedance $\zeta = O(a^{-1})$. We will discuss at
the end of Section \ref{S4} in more detail the question concerning the
compatibility of the assumption \eq{1.8}, namely $d\gg a$, and the
existence of the limits \eq{1.26b} and \eq{1.27}. It will be shown that
the assumption $d\gg a$ is compatible with the existence of 
the limit \eq{1.27} only if $\nu(y) $ is sufficiently small, and in this 
case the existence of the limit \eq{1.26b} is also compatible with the
assumption $d\gg a$. 

In Section \ref{S2} Theorems \ref{T2} and \ref{T3} are proved. In
Section \ref{S3} Theorem \ref{T3} and Lemmas 1, 2
are proved.
In Section \ref{S4} some examples are given, the significance of the
compatibility of the assumptions $d\gg a$ and \eq{1.21}, \eq{1.26b} --
\eq{1.27} is discussed, and a possible application of our results to
creating materials with a desired refraction coefficient is described.
\section{Proof of Theorem \ref{T2}}\label{S2}
Let us look for the solution to problem \eq{1.5} -- \eq{1.7} of the
form:
\begin{equation}
\label{E2.1}
u_M = u_0(x) + \sum^M_{m=1}\int_{S_m} G(x,s)\si_m(s)ds,
\end{equation}
where $G(x,y)$ is the resolvent kernel of $L_0$, see \eq{1.4}, and
$\si_m$ are arbitrary functions at the moment. For any $\si_m$ the
function \eq{2.1} solves equation \eq{1.5} and satisfies the radiation
condition \eq{1.6}. Since problem \eq{1.5} -- \eq{1.7} has at most one
solution, the function \eq{2.1} is the unique solution to \eq{1.5} --
\eq{1.7} provided that $\si_m$ are chosen so that the boundary
conditions \eq{1.7} are satisfied. Since ${\rm diam}\, D_m$, $1\leq m
\leq M$, are small, let us write \eq{2.1} as
\begin{equation}
\label{E2.2}
u_M = u_0(x) + \sum^M_{m=1} G(x,x_m)Q_m 
+\sum^M_{m=1}\int_{S_m}[G(x,s)-G(x,x_m)]\si_m(s)ds,
\end{equation}
where $x_m\in D_m$ is a point inside $D_m$ and
\begin{equation}
\label{E2.3}
Q_m:= \int_{S_m}\si_m(s) ds.
\end{equation}
The choice of $x_m\in D_m$ is arbitrary because ${\rm diam}\, D_m \leq
2a$ is small. We will prove that $Q_m \neq 0$, give an analytic formula 
for $Q_m$ (formula \eq{2.20} below), and
approximate the field $u_M$ in \eq{2.2} by the expression:
$$ 
u_M = u_0(x) + \sum^M_{m=1} G(x,x_m)Q_m.
$$
The error of this
approximate formula is of order $\max (\frac{a}{d}, ka)$, see estimate 
\eq{2.7} below. 
Therefore
this error tends to zero as $a\ra 0$ since $d = O(a^{1/3})$. 
Let us estimate the term 
\begin{equation}
\label{E2.4}
E_m:= \int_{S_m}[G(x,s)-G(x,x_m)]\si_m(s)ds.
\end{equation}
By the inequality  \eq{1.19} one gets
\begin{equation}
\label{E2.5}
\n E_m\n\leq c\Ba \frac{a}{d^2} + \frac{ka}{d}\Bz|Q_m|,
\quad \n x-x_m\n\geq d \gg a.
\end{equation}
We will prove below that $Q_m=O(a)$, see formula \eq{2.20},
and, since $ |G(x,x_m)|\leq cd^{-1}$ if $|x-x_m|\geq d>0$, one has: 
\begin{equation}
\label{E2.6}
\Bigl\n G(x,x_m)Q_m\Bigr\n = O\big(\frac a d\big).
\end{equation}
Let us prove that under our assumptions the term $E_m$
is much smaller than $O(\frac a d)$. Using again inequality \eq{1.19}, 
one gets:
$$|E_m|\leq c(ad^{-2}+kad^{-1})O(a).$$
Therefore,  
\begin{equation}
\label{E2.7}
|E_m|\leq O(\frac {a^2}{d^2}+ka\frac {a}{d})\ll O(\frac {a}{d}),
\end{equation}
because $ka\ll 1$ and $a\ll d$ by assumption. So, our claim is verified.
Moreover,  
$$ \sum^M_{m=1}|E_m|\ll  \sum^M_{m=1} |G(x,x_m)Q_m|$$ 
if $\n x-x_m\n \geq d \gg a$, because $M=O(\frac 1 a)$. 

To find $Q_m$, we use the boundary condition
\eq{1.7}. Let us write $u(x)$ in a neighborhood of $S_j$ as
\begin{equation}
\label{E2.8}
u_M(x) = u_e(x) + \int_{S_j}G(x,s)\si_j(s)ds,\quad \n x-x_j\n \leq 2a,
\end{equation}
where $u_e$ is the effective field acting on the $j-$th small particle
from outside:
\begin{equation}
\label{E2.9}u_e(x):= u_M(x)-\int_{S_j} G(x,s)\si_j(s)ds=
\sum_{m\neq j} G(x,x_m)Q_m +O(\frac a d).
\end{equation}
We neglect the error term  $O(\frac a d)$ in what follows.
From \eq{2.8} and \eq{1.7} one gets:
\begin{equation}
\label{E2.10}
0 = u_{eN}(s)-\zeta_j u_e(s) + \frac{A_j\si_j-\si_j}{2} - \zeta_j
T_j\si_j,\quad s\in S_j,
\end{equation}
where $u_{eN}(s)$ is the normal derivative of $u_e$ at the point $s\in 
S_j$. One can rewrite this equation as:
$$\si_j=A_j\si_j-2\zeta_j T_j\si_j-2\zeta_j u_e(s)+2 u_{eN}(s).$$
Here 
the operators $A_j$ and $T_j$ are defined as follows:
\begin{eqnarray}
\label{E2.11}
T_j\si_j&:= & \int_{S_j} G(s,t)\si_j(t)dt \simeq \int_{S_j }
\frac{\si_j(t)dt}{4\pi \n s-t\n}\,,
 \\
A_j\si_j & := & 2 \int_{S_j} \frac{\pa G(s,t)}{\pa N_s}\, \si_j(t)dt
\simeq 2 \int_{S_j} \frac{\pa }{\pa N_s} \: \frac{1}{4\pi \n s-t\n}\,
\si_j(t)dt := A \si_j,\hspace{1.5cm}\label{E2.12}
\end{eqnarray}
and we have used the following approximations:
\begin{align} G(x,y) &= g_0(x,y) [1 + O(\n x-y\n )], \quad\n x-y\n \ra
0;\quad g_0(x,y):= \frac{1}{4\pi \n x-y\n},\hspace{2cm}
\label{E2.13}\\
\frac{\pa G(x,y)}{\pa y_p} &= \frac{\pa g_0}{\pa y_p} \, \bka 1 + O\f \n
x-y\n^2\, \bigl\n \ln \n x-y\n\big\n\g \bkz, \quad\n x-y\n \ra 0.\label{E2.14}
 \end{align}
Note that (see \cite{R476}, p. 96, formula (7.21)):
\begin{equation}
	\label{E2.15}
	\int_{S_j} A\si ds = -\int_{S_j}\si_j ds
\end{equation}
Indeed,
	\[
	\int_{S_j} ds \int_{S_j} \frac{\pa}{\pa N_S} \frac{1}{2\pi | s-t|} 
\si(t)dt = \int_{S_j} dt \si_j(t) \int_{S_j}ds \frac{\pa}{\pa N_S} 
\frac{1}{2\pi | s-t|} = -\int_{S_j} \si_j(t)dt.
\]
The integral
	\[
	\int_{S_j} \frac{\pa}{\pa N_S} \frac{1}{2\pi | s-t|}ds=-1, 
\quad t \in S_j,
\]
is well known in potential theory for surfaces $S_j \in C^{1,\la}$.

Integrating \eq{2.10} over $S_j$, using formula \eq{2.15}, and the 
divergence theorem, one gets:
\begin{equation}
	\label{E2.16}
	Q_j = -\zeta_j u_e(x_j) | S_j| - \zeta_j \int_{S_j} ds \int_{S_j} 
\frac{\sigma_j(t)dt}{4\pi |s-t|} + \int_{D_j} \Da u_e dy.
\end{equation}
The function $u_e(y)$ is smooth,  so
\begin{equation}
	\label{E2.17}
	\int_{D_j} \Da u_e(y)dy= V_j\Da u_e (x_j) [1 +o(1)], \quad a\to 0,
\end{equation}
where $V_j = |D_j|$ is the volume of $D_j$ and we have used the smallness 
of the diameter of $D_j$, that is, the smallness of $a$.

Let us write
\begin{eqnarray}
	\nn \lefteqn{\int_{S_j} ds \int_{S_j} \frac{\si_j(t)dt}{4\pi 
|s-t|}  =  \int_{S_j} dt \, \si_j(t) \int_{S_j} \frac{ds}{4\pi |s-t|} }\\
	&  &= Q_j \frac{1}{S_j} \int_{S_j} dt \int_{S_j} \frac{ds}{4\pi |s-t|}
\label{E2.18}
	= \frac{Q_j J_j}{4 \pi|S_j|}, \quad J_j:=\int_{S_j} \int_{S_j} 
\frac{ds\, dt}{|s-t|}.
\end{eqnarray}
Here we approximated the continuous on $S_j$ function $\int_{S_j} \frac{ds}{|s-t|}$ by its mean value $\frac{1}{|S_j|}\int_{S_j} dt \int_{S_j} \frac{ds}{|s-t|}$.

If $S_j$ is a sphere of radius $a$, then
\begin{equation}
\label{E2.19}
\int_{\n s\n = a} \frac{ds}{\n s-t\n} = 4\pi a,\quad \n t\n = a,
\end{equation}
so in this case equation \eq{2.18} is exact. 

From \eq{2.16}  --
\eq{2.18} one finds a formula for $Q_j$:
\begin{equation}
\label{E2.20}
Q_j = - \frac{\zeta_j \n S_j\n}{1 + \frac{\zeta_j J_j}{4\pi \n S_j\n}}\, 
u_e(x_j).
\end{equation}
We neglected the term $V_j \, \Da u_e(x_j) = O(a^3)$ which is much
smaller than $\n \zeta_j\n\: \n S_j\n = O(a)$ as $a \ra 0$, because
$\n S_j\n = O(a^2)$ and $\n \zeta_j\n = O(\frac{1}{a})$. The quantity
$J_j = O(a^3)$. Therefore $\frac{\zeta_j J_j}{4\pi \n S_j\n} = O(1)$.
We choose
\begin{equation}
\label{E2.21}
\zeta_j = \frac{H(x_j)}{a}\,,
\end{equation}
where $H(x)$ is a continuous function in $D$, which we can choose as we
wish subject to the condition $\Im H\leq 0$, because $\Im \zeta_j \leq 0.$

If the small particles are all of the same shape and size then $\n
S_j\n = c_1 a^2$, and $J_j = c_2 a^3$, where $c_1,c_2 > 0$ are constants
 independent of $j$, $1\leq j \leq M$, and then
\begin{equation}
\label{E2.22}
\frac{\zeta_j J_j}{4\pi \n S_j\n} = \frac{H(x_j)c_2}{4\pi c_1}:= h(x_j),
\end{equation}
and
\begin{equation}
\label{E2.23}
\zeta_j \n S_j\n = H(x_j) c_1 a.
\end{equation}
Formulas \eq{2.2}, \eq{2.20}, \eq{2.22} and \eq{2.23} imply:
\begin{equation}
\label{E2.24}
u_M(x) = u_0(x)-\sum^M_{m=1}G(x,x_m)\, \frac{4\pi c_1^2 c_2^{-1}h(x_m)
a}{1 + h(x_m)}\, u_M(x_m),
\end{equation}
where $\n x-x_m\n \geq d \gg a$, and we replaced $u_e(x_m)$ by $u_M(x_m)$
because their difference (see \eq{2.9}) is of order $O(\frac{a}{d})\ll
1$. Indeed
\begin{equation}
\label{E2.25}
\n u_M(x)-u_e(x)\n \leq \int_{S_j} \n G(x,s)\n\: \n \si_j(s)\n ds \leq
\frac{c}{d}\, \n Q_j\n \leq \s{c}\, \frac{a}{d}\,,\quad \n x-x_j\n \geq
d \gg a,
\end{equation}
where $c,\s{c} > 0$ are some constants independent of $a$. 

If the
assumption \eq{1.21} holds, then
\begin{equation}
\label{E2.26}
\lim_{a\ra 0} \sum^M_{m=1} G(x,x_m)\, \frac{4\pi c_1^2
c_2^{-1} h(x_m)}{1 + h(x_m)}\, u_M(x_m) a
= \int_D G(x,y) \, \frac{4\pi c_1^2 c_2^{-1}h(y)}{1 +
h(y)}\, u(y) N(y) dy.
\end{equation}
To pass to the limit in \eq{2.26} one can use the following lemma.
\begin{lem}\label{L3}
Assume that $x_m\in D_m$, ${\rm diam}\, D_m\leq 2a$, $f$ is a continuous
function in $D$ with a possible exception of a point $y_0$ in a 
neighborhood of which
it is absolutely integrable, for example, it
admits an estimate $\n f(y)\n \leq \frac{c}{\n y-y_0\n^b}$, $b < 3$, and
assume that
\begin{equation}
\label{E2.27}
\lim_{a\ra 0} a\sum_{D_m\subset \s{D}} 1 = \int_{\s{D}} N(x)dx\quad \fa
\s{D}\subset D
\end{equation}
for any subdomain $\s{D}\subset D$, where $N(x)$ is a continuous
function. Then there exists the limit
\begin{equation}
\label{E2.28}
\lim_{a\ra 0}\sum^M_{m=1}f(y_m)a = \int_D f(y) N(y)dy.
\end{equation}
\end{lem}
\begin{rem}\label{R1}
{\rm In our case $f(y) = G(x,y)\, \frac{4\pi c_1^2 h(y)}{c_2 [1 +
h(y)]}\, u_M(y)$ and \eq{2.27} is the assumption \eq{1.21}.
}
\end{rem}
{\bf Proof of Lemma \ref{L3}} Let $D = \bigcup^P_{p=1} \ov{\Da}_p$,
where   $\Da_p $ and $\Da_q$ do not 
intersect each other, $\ov{\Da}_p$ is
the closure of the domain $\Da_p$, and 
$\lim_{P\ra \ue}\,\max_{p} {\rm diam}\,
\Da_p = 0$. Choose any point $y^{(p)}\in \Da_p$ and note that
\begin{equation}
\label{E2.29}
\sup_{y_m\in D_m,\, D_m\subset \Da_p}\, \n f(y^{(p)}) - f(y_m)\n < \ve_p
\ra 0\quad \mbox{as diam}\, \Da_p \ra 0.
\end{equation}
Therefore 
\begin{eqnarray}
\nn \lim_{a\ra 0}\sum^M_{m=1} f(y_m)a & = & \lim_{a\ra 0 }\sum^P_{p=1} a
\sum_{D_m\subset \Da_p} f(y_m) = \sum^P_{p=1}[f(y^{(p)}) + O(\ve_p)]
\cdot \lim_{a\ra 0} a \sum_{D_m\subset \Da_p}1 \\
\nn & = & \sum^P_{p=1}[f(y^{(p)}) + O(\ve_p)] \int_{\Da_p}N(y)dy \\ 
&=& \sum^P_{p=1}
[f(y^{(p)}) + O(\ve_p)] \cdot [N(y^{(p)}) + O(\ve'_p)]\:\n \Da_p\n,\label{E2.30}
\end{eqnarray}
where $\lim_{P\ra \ue}\max_p\n \ve'_p\n = 0$. Let $P \ra \ue$ in 
\eq{2.30}.
Then
\begin{equation}
\label{E2.31}
\lim_{P\ra \ue} \sum^P_{p=1} [f(y^{(p)}) + O(\ve_p)]\: [N(y^{(p)}) +
O(\ve c'_p)]\, \n \Da_p\n = \int_D f(y)N(y)dy.
\end{equation}
In the above argument we assumed that $f$ is continuous in $D$. If $f$
has an integrable singularity at a point $x_0$, then we choose a ball
$B(x_0,\da_\ve)$ centered at $x_0$ of radius $\da_\ve$ such that
$\sup_{0 < \da < \da_\ve} \int_{B(x_0,\da)}\n f(y)\n dy < \ve$, where
$\ve > 0$ is an arbitrary small fixed number. Then 
$$\sup_{0<\da <\da_\ve}\int_{B(x_0,\da)} \n f(y)\n\: \n N(y)\n\, dy < 
c\ve,$$ 
where $c= \max_{y\in D}\, \n N(y)\n > 0$ is a constant independent of $\ve$.
Now we apply the above argument to the region $D\bs B(x_0,\da)$, where
$f$ is continuous and get:
\begin{equation}
\label{E2.32}
\lim_{a\ra 0}\sum^M_{\substack{m=1\\ y_m\not\in B(x_0,\da)}} f(y_m)a =
\int_{D\bs B(x_0,\da)} f(y)N(y)dy.
\end{equation}
The left side of \eq{2.28} in the case of $f$ having integrable
singularity at the point $x_0$ and continuous in $D\bs x_0$ is
understood as the limit of the expression on the left of \eq{2.32} as
$\da \ra 0$. This yields \eq{2.28}. Lemma \ref{L3} is proved.\qed
\vpar
Passing to the limit $M\ra \ue$, or $a\ra 0$, in equation \eq{2.24} and
using Lemma \ref{L3}, one gets
\begin{eqnarray*}
u(x) &= & u_0(x) - \int_D G(x,y)\, p(y) \, u(y)dy, \\
p(x) & = & \frac{4\pi c^2}{c^2}\: \frac{h(y)N(y)}{1 + h(y)}.
\end{eqnarray*}
Applying the operator $L_0 = \na^2 + k^2 - q_0(x)$ to this equation and
using \eq{1.3} and \eq{1.4}, one obtains equation \eq{1.23}. Formulas
\eq{1.25} and \eq{1.26a} follow from the above equation and from formula
\eq{1.12}.\vpar This concludes the proof of Theorem \ref{T2}.\qed
\begin{rem}\label{R2}
{\rm It is possible (and not difficult) to generalize Theorem \ref{T2}
to the case of particles with different shapes. Since this does not lead
to an essentially new result, we do not go into detail. In [8] one
can find analytical formulas for the $S-$matrix for wave scattering by 
small bodies of arbitrary shapes. 
}
\end{rem}
{\bf Proof of Theorem \ref{T3}}. 

Now we assume $\zeta_m=0$, $1\leq m
\leq M$, which means that all the small particles are acoustically hard. 
In this case equation \eq{2.10} takes the form
\begin{equation}
\label{E2.33}
\si_j = A_j\si_j + 2u_{e_N}(s),\quad s\in S_j,\; 1 \leq j \leq M,
\end{equation} 
where 
\begin{equation}
\label{E2.34}
u_e(x):= u_0(x) + \sum_{m\neq j}\int_{S_m} G(x,s)\si_m(s)ds.
\end{equation}
We cannot use approximation \eq{2.2} because the quantity $Q_m$ now is
of the same order of magnitude as the integral $\int_{S_m}
[G(x,s)-G(x,x_m)]\, \si_m(s)ds$, or even smaller than this integral. 
This is established below. While
under the assumptions of Theorem \ref{T2} we had $Q_m = O(a)$, now,
under the assumptions of Theorem \ref{T3}, we have $Q_m = O(ka^2 a^3)$,
which is a much smaller quantity than $O(a)$ because $ka \ll 1$.
To estimate the order of magnitude of $Q_m$, we integrate \eq{2.33} over
$S_j$ and use \eq{2.15}. The result is:
\begin{equation}
\label{E2.35}
Q_j = \int_{S_j} u_{e_N} ds = \int_{D_j}\Da u_e \, dx \simeq \Da u_e(x_j)V_j,
\end{equation}
where $V_j$ is the volume of $D_j$, and we have used the assumption $d\gg
a$. This assumption allows one to claim that $u_e(x)$  is practically
constant in the domain $D_j$ in the absence of $j$-th particle.
Differentiation with respect to $x$ brings a factor $k$. Since we assume
that $k > 0$ is fixed, this factor is not important for our argument,
but to make the dimensionality of the term $V_j \Da u_e$ clear, we may
write $V_j \Da u_e = O(k^2 a^3)$. This quantity has dimensionality of
length since $ka$ is dimensionless.

We now prove that the term $E_m:=\int_{S_m}[G(x,s)-G(x,x_m)] \si_m(s)ds$,
which was neglected under the assumptions of Theorem \ref{T2}, because it
was much smaller than $\n G(x,x_m)Q_m\n$, is now, under the assumption $\zeta_m
= 0$, $1 \leq m \leq M$, of the same order of magnitude as
$\n G(x,x_m)Q_m\n$, namely $O(k^2 a^3 d^{-1})$, or even larger. We have
\begin{eqnarray}
\nn\lefteqn{ \int_{S_m} [G(x,s)-G(x,x_m)]\si_m(s)ds
}\\
\label{E2.36}
&& = \int_{S_m} \na_y G(x,a)\rain{y=x_m}\cdot (s-x_m)\, \si_m(s)ds,\quad
\n x-x_m\n \geq d\gg a,
\end{eqnarray}
where we have used the assumption $\n x-x_m\n \gg a$ and kept
the main term in the Taylor's expansion of the function $G(x,s)-G(x,x_m)$.

Recall, that
\begin{equation}
\label{E2.37}
\int_{S_m}(s-x_m)_p\, \si_m(s)ds = -V_m\, \be^{(m)}_{pj}\,
\frac{\pa u_e(y)}{\pa y_j}
\Bigr\n_{y=x_m},
\end{equation}
where $\be_{pj}^{(m)}$ is the magnetic polarizability tensor defined in
\cite{R476}, (p.55, formulas (5.13)-(5.15) and p.62, formula (5.62)), and 
$(s-x_m)_p$ is the $p$-th component of the vector
$s-x_m$. Namely, if
\begin{equation}
\label{E2.38}
\si = A\si - 2N_j,
\end{equation}
then
\begin{equation}
\label{E2.39}
\int_S s_p \, \si(s)ds = V\be_{pj}\,,
\end{equation}
where $V$ is the volume of the body with boundary $S$, $N_j$ is the
$j$-th component of the exterior unit normal $N$ to $S$, the role of 
the point $x_m$
from equation \eq{2.37} is played by the origin, which is located inside 
$S$,
and the role of $S_m$ is played by $S$. Equation \eq{2.33} with $j=m$
can be written as
\begin{equation}
\label{E2.40}
\si_m = A_m\si_m - 2N_j\Ba {-} \frac{\pa u_e(x_m)}{\pa y_j}\Bz.
\end{equation}
Compare \eq{2.40} and \eq{2.38} and get \eq{2.37}.

Formulas for the tensor $\be_{pj}=\al_{pj}(\gm)\Bigr\n_{\gm=-1}$ for
bodies of arbitrary shapes are derived in \cite{R476}, p.55,  formula 
(5.15),
so one may consider the tensor $\be_{pj} $ known for bodies of arbitrary
shapes. The parameter $\gamma=\frac 
{\epsilon_i-\epsilon_e}{\epsilon_i+\epsilon_e}$, where $\epsilon_i$
is the dielectric permittivity of the body and $\epsilon_e$
is the dielectric permittivity of the surrounding medium. The case
$\gamma=-1$ occurs when $\epsilon_i=0$. This is the case, for example,
in the problem of calculation the magnetic dipole moment of a 
superconductor placed in a homogeneous magnetic field: in the 
superconductor the magnetic induction vector $B=0$, which means
that the magnetic permeability $\mu_i$ of such body is zero,
$\mu_i=0$,  see \cite{L}. That is why the tensor $\beta_{pj}$
is called magnetic polarizability tensor in \cite{R476}.

From \eq{2.36} and \eq{2.37} it follows that
\begin{equation}
\label{E2.41}
\int_{S_m}\bka G(x,s)-G(x,x_m)\bkz\, \si_m(s)ds = - \frac{\pa
G(x,y)}{\pa y_p}\Bigr\n_{y=x_m}\,\frac{\pa u_e(y)}{\pa y_j}
\Bigr\n_{y=x_m}\, V_m \be^{(m)}_{pj},
\end{equation}
where one sums up over the repeated indices $p,j$, but nor over $m$.
The quantity on the right of \eq{2.41} is of the order $O(k^2 a^3
d^{-1})$ if $kd\geq 1$, that is, of the same order as $\n G(x,x_m)Q_m\n$, 
provided
that  $\n x-x_m\n\geq d\gg a$, and
it is of the order $O(ka^3d^{-2})$ if $kd<1$. 
Indeed,  $\be^{(m)}_{pj}=O(1)$, 
$V_m =O(a^3)$, and 
$\big|{\na_y G(x,y)}\big | \leq c \max \f \frac{k}{d}, 
\frac{1}{d^2}\g$.
\par
Let us prove the estimate 
$$\big| \na_y G(x,y)\big | \leq c \max \f\frac{k}{d},
\frac{1}{d^2}\g \quad \mbox{\quad for \quad } |x-y| \geq d\gg a,$$ 
where $c>0$ is a constant independent of $d$. 

We have
	\[G(x,y)=g(x,y)-\int_D g(x,z) q_0(z)G(z,y)dz:=g-TG,
\]
where $T$ is compact as an operator in $L^p(D),p\geq 1$ under our 
assumptions, namely, $D \subset R^3$ is a bounded domain, $q_0(x)$ is a 
bounded piecewise-continuous function. From this equation we get 
$$\na_y G(x,y)=\na_y g(x,y)-T\na_y G.$$
\par
Clearly, 
$$\na_y g=g(ik-\frac{1}{|x-y|}) \frac{y-x}{|x-y|},$$ 
so
	\[ | \na_y g(x,y)| \leq 2 \max \f \frac{k}{2\pi d},
\frac{1}{4\pi d^2} \g = \frac{1}{2\pi}\max \f\frac{k}{d},\frac{1}{d^2}\g, 
\quad |x-y|\geq d >0.
\]
Thus
	\[ | \na_y G(x,y)| \leq |\na_y g(x,y)| 
\big [ 1+c\int_D \frac{1}{|x-z|}|\na_y G(z,y)|dz \frac{|x-y|^2}{|ik|x-y|-1|} 
\big ] := |\na_y g| (1+cI),
\]
where
	\[ I:=\int_D |\na_y G(z,x)| \frac{dz}{|x-z|} \frac{|x-y|^2}
{\sqrt{1+k^2|x-y|^2}}\leq c \int_D \frac{dz}{|z-y|^2 |x-z|} \frac{|x-y|^2}
{\sqrt{1+k^2d^2}}.
\]
One has
	\[ \int_D \frac{dz}{|z-y|^2 |x-z|} \leq c \big | \ln |x-y|\big|,
\]
and 
$$\sup_{x,y\in D} |\ln |x-y| \big | |x-y|^2 \leq c,$$ 
where $c=c(D)$ is a 
constant. Therefore 
$$I\leq \frac{c}{\sqrt{1+k^2d^2}}\leq c,$$ 
and
	\[|\na_y G(x,y)| \leq c \max \f \frac{k}{d},\frac{1}{d^2}\g \frac{1}{\sqrt{1+k^2d^2}}\leq c \max \f \frac{k}{d},\frac{1}{d^2}\g,
\]
as claimed.
\par
If $\frac{k}{d} \geq \frac{1}{d^2}$, i.e. $kd\geq 1$, then $|\na_y G(x,y)| 
\leq c \frac{k}{d},\,\,\, |x-y|\geq d >0$.
\par
If $\frac{k}{d}<\frac{1}{d^2}$, i.e. $kd<1$, then $| \na_y G(x,y)| 
\leq \frac{c}{d^2}, \,\,\,|x-y|\geq d>0$.
\par
Therefore, the right side of \eq{2.41} is 
$O\f \frac{k^2a^3}{d}\g$ if $kd\geq 1$, 
in which case it is of the same order as the term $G(x,x_m)Q_m$. If $kd<1$, 
then the right side of \eq{2.41} is 
$O\f \frac{ka^3}{d^2}\g$, in which case it may become larger than the term 
$G(x,x_m)Q_m$ because the ratio $\frac{ka^3}{d^2}/ \frac{k^2a^3}{d} = 
\frac{1}{kd}>1$ provided that $kd<1$.
\par
Writing the
field \eq{2.1} in the form
\begin{equation}
\label{E2.42}
u_M(x) = u_0(x) + \sum^M_{m=1} G(x,x_m)Q_m + \sum^M_{m=1}\int_{S_m} \bka
G(x,s)-G(x,x_m)\bkz\, \si_m(s)ds
\end{equation}
and using formulas \eq{2.35} and \eq{2.41}, one gets:
\begin{equation}
\label{E2.43}
u_M(x)=u_0(x) + \sum^M_{m=1}G(x,x_m)\Da u_e(x_m)\, V_m - \sum^M_{m=1}
\frac{\pa G(x,x_m)}{\pa y_p}\: \frac{\pa u_e(x_m)}{\pa y_j}\, V_m \be^{(m)}_{pj}(x_m),
\end{equation}
and over the repeated indices $p,j$ one sums up.

Let $a\ra 0$, $M\ra \ue$. We want to give sufficient conditions for
passing to this limit in \eq{2.43}.
\begin{lem}\label{L4}
Assume that for any subdomain $\s{D}\subset D$ the following limits 
exist:
\begin{eqnarray}
\lim_{a\ra 0}\sum_{D_m\subset \s{D}} V_m \be^{(m)}_{pj}(x_m)& =&
\int_{\s{D}}\be_{pj}(y)\, \nu (y)dy,\label{E2.44} \\
\label{E2.45} \lim_{a\ra 0}\sum_{D_m\subset \s{D}} V_m & = &
\int_{\s{D}} \nu(y)dy.
\end{eqnarray}
Then the limiting form of equation \eq{2.43} is:
\begin{eqnarray}
\nn \U(x) & =& u_0(x) + \int_D G(x,y)\, \Da \U(y)\, \nu(y) dy \\
&& \mbox{} - \int_D \frac{\pa G(x,y)}{\pa y_p}\: \frac{\pa \U(y)}{\pa
y_j}\, \be_{pj}(y)\, \nu(y)dy,\label{E2.46}
\end{eqnarray}
where one sums up over the repeated indices $p,j$.
\end{lem}
\begin{rem}\label{R3}
{\rm If one assumes that $\nu(y)$ vanishes near the boundary $S$ of $D$
and integrates  the last integral in \eq{2.46} by parts, one gets
\begin{equation}
\U(x) =  u_0(x) + \int_D G(x,y) \Big\{ \Da \U(y)\nu(y) 
 + \sum^3_{p,j=1}
\frac{\pa}{\pa y_p} \Ba \frac{\pa \U(y)}{\pa y_j}\, \be_{pj}(y)\,
\nu(y)\Bz\Big\}dy.\label{E2.47}
\end{equation}
Applying the operator $L_0$ to both sides of \eq{2.47} and using
\eq{1.4} one gets:
\renewcommand{\theequation}{\thesection.\arabic{equation}'}
\addtocounter{equation}{-1}
\begin{equation}
\label{E2.47'}
L_0\U + \nu(y) \Da \U(x) + \sum^3_{p,j=1} \frac{\pa}{\pa y_p}\, \Ba
\frac{\pa \U(y)}{\pa y_j}\, \be_{pj}(y)\, \nu(y)\Bz = 0.
\end{equation}
\renewcommand{\theequation}{\thesection.\arabic{equation}}
}\end{rem}
\begin{rem}\label{R4}
{\rm If all the  small particles are identical, then $V_m=c_3a^3$, where
the positive constant $c_3$ does not depend on $m$, and
$\be^{(m)}_{pj}=\be_{pj}$. Then
\begin{equation}
\label{E2.48}
\lim_{a\ra 0}\sum_{D_m\subset \s{D}} V_m = \lim_{a\ra 0}\Bka c_1 a^3
\sum_{D_m\subset \sD} 1 \Bkz = \lim_{a\ra 0}[c_1 a^3 N(\sD)],
\end{equation}
where $N(\sD)$ is the number of small particles in the domain $\s{D}$. For 
the
limit \eq{2.48} to exist it is sufficient that
\begin{equation}
\label{E2.49}
N(\sD) = \frac{\int_{\sD} \nu(y)dy}{c_1 a^3}\,,
\end{equation}
where $\nu(y)\geq 0$ is a continuous function, 
and the limit in  \eq{2.48} is equal to $\int_{\sD} \nu(y)dy$.

One can write \eq{2.49} as
\begin{equation}
\label{E2.50}
N(y)dy = \frac{\nu(y)}{c_1 a^3}\, dy.
\end{equation}
In contrast to Theorem 2, where $M=O(\frac 1 a)$, we now have 
$M=O(\frac 1 {a^3})$.

Similarly,
\begin{equation}
\label{E2.51}
\lim_{a\ra 0}\sum_{D_m\subset \sD}V_m \be^{(m)}_{pj}(x_j) = \lim_{a\ra
0} [c_1 a^3 \, \be_{pj}\, N(\sD)] = \be_{pj}\int_\sD \nu(y) dy.
\end{equation}
}\end{rem}
We gave in this Remark some practically realizable sufficient conditions 
for the
existence of the limits \eq{2.44} and \eq{2.45}.

Let us verify
that if the limits \eq{2.44} -- \eq{2.45} exist, then the limit of the
right side of equation \eq{2.43} exists, and, denoting this limit by
$$\U(x)=\lim_{a\ra 0}u_M(x),$$ 
one obtains the limiting form of equation
\eq{2.43}:
\begin{eqnarray}
\nn \U(x) &= & u_0(x) + \int_D G(x,y)\, \Da \U(y)\, \nu(y)dy \\
&& \mbox{} - \int_D \sum^3_{p,j=1} \frac{\pa G(x,y)}{\pa y_p}\:
\frac{\pa \U(y)}{\pa y_j}\, \be_{pj}(y)\, \nu(y)dy,\label{E2.52}
\end{eqnarray}
which is equation \eq{2.46}. 

We took into account that
$$\lim_{a\ra 0}u_e(x) = \U(x).$$
This is so because, as $a\to 0$, the input of a single particle into
the field $ \U(x)$ tends to zero.

Let us verify the existence of the limit of the right side of equation
\eq{2.43}.
We use, as in the proof of Lemma \ref{L3}, a representation of $D$ of
the form $D = \bigcup^P_{p=1} \ov{\Da}_p$, and assume that
\begin{equation}
\label{E2.53}
\lim_{P\ra \ue}\max_{1\leq p \leq P}\,{\rm diam}\, \Da_p = 0.
\end{equation}
Then
\begin{eqnarray}
\nn \lefteqn{\lim_{a\ra 0}\sum^M_{m=1}G(x,y_m)\Da u_e(x_m)V_m
}\\
\nn & & = \sum^P_{p=1}\lim_{a\ra 0} \sum_{D_m\subset \Da_p} G(x,x_m)\Da
u_e(x_m) V_m\\
\nn & & = \sum^P_{p=1} G(x,y^{(p)})\, \Da u_e(y^{(p)})(1 + \ve_p)\,
\lim_{a\ra 0}\sum_{D_m\subset \Da_p} V_m\\
\label{E2.54}& & = \sum^P_{p=1} G(x,y^{(p)})\, \Da u_e (y^{(p)})(1 +
\ve_p)\, \nu^2 (y^{(p)})(1 + \ve'_p)\n \Da_p\n,
\end{eqnarray}
where 
\begin{equation}
\label{E2.55}
\lim_{P\ra \ue}\max_p(\n \ve_p\n + \n \ve'_p\n) = 0.
\end{equation}
Let $P \ra \ue$ in \eq{2.54} and use \eq{2.55} to get
\begin{eqnarray}
\nn && \lim_{P\ra \ue} \sum^P_{p=1} G(x,y^{(p)})\, \Da u_e(y^{(p)})\,
\nu(y^{(p)})\, \n \Da_p\n\, (1 + \ve_p + \ve'_p + \ve_p\ve'_p)\hspace{1.5cm}
\\
\label{E2.56}&& \hph{\lim_{P\ra \ue} \sum^P_{p=1}} = 
\int_D G(x,y)\, \Da \U(y)\, \nu(y)dy.
\end{eqnarray}
We have replaced $u_e(y^{(p)})$ in the limit $P\ra \ue$ by $\U(y)$,
because
\begin{equation}
\label{E2.57}
\U(y) - u_e(y) = \int_{S_m}G(s,t)\, \si(t)dt = o(1)\quad \mbox{as } a\ra 0.
\end{equation}
From \eq{2.56} and \eq{2.54} one gets:
\begin{equation}
\label{E2.58}
\lim_{a\ra 0}\sum^M_{m=1}G(x,x_m)\, \Da u_e(x_m) V_m = \int_D G(x,y)\,
\Da \U(y)\, \nu(y)dy.
\end{equation}
The singular points $x = y\in D$ of $G(x,y)$ are treated as in the proof
of Theorem \ref{T2}. 

The function $\n G(x,y)\n \leq c\n x-y\n^{-1}$ as
$\n x-y\n \ra 0$, so $\n G(x,y)\n\in L^1(D)$ as a function of $y$ for
any $x\in D$.

Similar arguments, applied to the last sum in \eq{2.43}, lead to the
formula
\begin{equation}
\label{E2.59}
\lim_{a\ra 0}\sum^M_{m=1}\frac{\pa G(x,x_m)}{\pa y_p}\: \frac{\pa
u_e(x_m)}{\pa y_j}\, V_m\, \be^{(m)}_{pj}(x_m) = \int_D \frac{\pa
G(x,y)}{\pa y_p}\: \frac{\pa \U}{\pa y_j}\, \be_{pj}(y)\, \nu(y)dy,
\end{equation}
where one sums up over the repeated indices $p,j$.

Theorem \ref{T3} is proved. \qed
\vpn

In Section \ref{S4} we discuss the compatibility of the condition $d \gg 
a$ and the existence of the limit \eq{1.27}.
\section{Auxiliary results}\label{S3}
In this Section we prove Theorem \ref{T1} and Lemmas \ref{L1}, \ref{L2}.
\vpn
{\bf Proof of Lemma \ref{L1}} Let us start with the following observations:
\begin{eqnarray}
	\label{E3.1}
	&&\big| |t-y| - |x-y| \big| \leq |t-y-(x-y)| = |t-x|\leq a,\\
	\label{E3.2}
	&&\sup_{-a\leq s \leq a} | e^{is} -1| \leq a,\\
	\label{E3.3}
	&&\big | e^{ik|t-y|} - e^{ik|x-y|} \big| = \big| e^{ik(|t-y|-|x-y|)} -1\big| \leq ka,
\end{eqnarray}
where the last inequality follows from \eq{3.2}.

One has
\begin{eqnarray}
\nn\big | g(t,y) - g(x,y)\big | &=& \frac{\big | |x-y| e^{ik|t-y|} - 
|t-y| e^{ik|x-y|} \big|}{|x-y||t-y|} \leq \frac{\big | |x-y| - |t-y| \big|}
{|x-y| |t-y|} +\\[2ex]
\label{E3.4} &&+ \frac{|t-y| \big | e^{ik|t-y|} - e^{ik|x-y|}}{|x-y||t-y|}\\
\nn &\leq &\frac{a}{|x-y||t-y|} + \frac{ka}{|x-y|}\leq \frac{a}
{d^2(1-\frac{a}{d})} + \frac{ka}{d}\leq O\f \frac{a}{d^2}\g + \frac{ka}{d}.
\end{eqnarray}
Lemma \ref{L1} is proved. \qed
\vpn
{\bf Proof of Lemma \ref{L2}} Let us start with the equation:
\begin{equation}
	\label{E3.5}
	G(x,y)=g(x,y)-\int_D g(x,z) q_0(z)G(z,y)dz,
\end{equation}
where $q_0$ is defined in \eq{1.2}. From \eq{3.5} one gets:
\begin{eqnarray}
	\nn \big| G(t,y) - G(x,y)\big| &\leq& \big | g(t,y) - g(x,y)\big| + 
\big| \int_D [g(t,z) - g(x,z)] q_0(z)G(z,y)dz\big|\\
&\leq& O\f \frac{a}{d^2}) + \frac{ka}{d}+
c\int_D|g(t,z)-g(x,z)|\frac{dz}{|z-y|}	.\label{3.6}
\end{eqnarray}
Here we have used Lemma \ref{L1} and the estimates
\begin{equation}
	\label{E3.7}
	\sup_{z\in D} |q_0(z)| \leq c_4, \quad |G(z,y)| \leq c_5|z-y|^{-1},
\end{equation}
where $c_4,c_5>0$ are some constants.

Let us estimate the integral
\begin{eqnarray}
	\nn I & := & \int_D |g(t,z) - g(x,z)| \frac{dz}{|z-y|} \\
\nn 	&=& \int_{|x-z|\geq \frac{d}{4}, z\in D} \frac{|g(t,z)-g(x,z)| 
dz}{|z-y|} + \int_{|x-z|\geq \frac{d}{4}, z\in D} \frac{|g(t,z)-g(x,z)| 
dz}{|z-y|}
	\\
\label{E3.8}	& := & I_1 + I_2. 
\end{eqnarray}
By Lemma \ref{L1}, which is applied to $I_1$ with $d$ replaced by $\frac{d}{4}$, one gets
\begin{equation}
	\label{E3.9}
	I_1 \leq c\f \frac{a}{d^2} + \frac{ka}{d}\g \int_{|x-z|\geq \frac{d}{4}} \frac{dz}{|z-y|} \leq c_1 \f \frac{a}{d^2} + \frac{ka}{d}\g.
\end{equation}
Here and below we do not write $z \in D$ under the integration sign to simplify the notations.

Let us estimate $I_2$:
\begin{equation}
	\label{E3.10}
	I_2 \leq \frac{1}{4\pi} \int_{|x-z| \leq \frac{d}{4}} \frac{dz \big| e^{ik|t-z|} |x-z| - e^{ik|x-z|} |t-z|\big|}{|z-y||t-z||x-z|}.
\end{equation}
One has
\begin{eqnarray}
	\nn \big| e^{ik|t-z|} |x-z| - |t-z| e^{ik|x-z|} \big |& \leq & \big| |x-z| - |t-z| \big |+ |t-z| \big | e^{ik|t-z|} - e^{ik|x-z|} \big| \\
	\label {E3.11} & \leq & |x-t| + |t-z| k \big | |t-z| - |x-z| \big| \\
\nn	&\leq& |x-t| + k|t-z| |t-x|.
\end{eqnarray}
Thus, with $|x-y| \geq d \gg a$ and $|t-x| \leq a$, one has:
\begin{eqnarray}
 \nn I_2 & \leq & \frac{1}{4\pi} \int_{|x-z| \leq \frac{d}{4}} 
\frac{dz (|x-t| + k|t-z||t-x|)}{|z-y||t-z||x-z|}\\
 \nn & \leq & \frac{|t-x|}{4\pi} \Big (\int_{|x-z| \leq \frac{d}{4}} 
\frac{dz}{|z-y||t-z||x-z|} + k\int_{|x-z| \leq \frac{d}{4}} 
\frac{dz}{|z-y||x-z|}\Big ) \\
 &\leq& ca \f \frac{1}{d^2} + \frac{k}{d}\g.\label{E3.12}
\end{eqnarray}
From \eq{3.9} and \eq{3.12} the estimate \eq{1.19} follows. Lemma \ref{L2} is proved. \qed
\vpn
{\bf Proof of Theorem \ref{T1}} Let us first prove that if conditions 
\eq{1.14} hold, then problem \eq{1.5} -- \eq{1.7} has at most one 
solution. It is sufficient to prove that the homogeneous problem
\begin{eqnarray}
	\label{E3.13} (\na^2 + k^2 - q_0)u =0\text{ in } 
\R^3 \setminus \bigcup_{m=1}^M D_m,\\
	\label{E3.14} \frac{\pa u}{\pa r} - iku = o(\frac{1}{r}), \quad 
u=O(\frac{1}{r}), \quad r=|x| \to \infty,\\
	u_N=\zeta_m u \text{ on } S_m, \quad 1\leq m \leq M, \label{E3.15}
\end{eqnarray}
has only the trivial solution if conditions \eq{1.14} hold. 

Taking complex conjugate of \eq{3.13} -- \eq{3.15} one gets:
\begin{eqnarray}
	\label{E3.16} \f \na^2 + k^2-\overline{q}_0(x) \g \overline{u} =0 
\text{ in } \R^3 \setminus \bigcup_{m=1}^M D_m,\\
	\label{E3.17} \frac{\pa \ov{u}}{\pa r} + ik\ov{u} =o\f 
\frac{1}{r}\g, \quad \ov{u} = O\f \frac{1}{r}\g, \ r=|x| \to \infty,\\
	\label{E3.18} \ov{u}_N = \ov{\zeta}_m \ov{u} \text{ on } S_m, 
\quad 1\leq m \leq M.
\end{eqnarray}
Multiply \eq{3.13} by $\ov{u}$, \eq{3.16} by $u$, subtract
from the first equation the second one, and integrate 
over the region  $(\R^3\setminus \bigcup_{m=1}^M D_m) \cap B_R := D_R,$ 
where $B_R$ is the ball centered at the origin of radius $R$. 
Using Green's formula, one gets:
\begin{eqnarray}
	\nn 0 & = & \int_{D_R} [\ov{u} \na^2 u - u \na^2 \ov{u} - 
(q_0-\ov{q}_0) |u|^2] dx \\
\nn	 &=& -2i\int_{D_R} \Im {q}_0(x) |u|^2 dx + \int_{|x|=R} \big( \ov{u} \frac{\pa u}{\pa r} - u \frac{\pa \ov{u}}{\pa r}\big) ds \\
\label {E3.19}&&- \sum_{m=1}^M \int_{S_m} \big ( \ov{u} \frac{\pa u}{\pa N} - u \frac{\pa \ov{u}}{\pa N}\big ) ds.
\end{eqnarray}
Using \eq{3.17} and \eq{3.18} one rewrites \eq{3.19} as follows:
\begin{equation}
	\label{E3.20}
	0=-2i \int_{D_R} \Im q_0(x) |u|^2 dx 
+2ik\int_{|x|=R}|u|^2 ds+o(1) 
- 2i\sum_{m=1}^M \int_{S_m} \Im \xi_m |u|^2 ds.
\end{equation}
Letting $R\to \infty$,  taking into account that $q_0(x)=0$ in 
$D'=\R^3\setminus D$, and one gets:
\begin{equation}
	\label{E3.21}
	0\leq \int_{D\setminus \bigcup_{m=1} ^M D_m} \Im q_0(x) |u|^2 dx + 
\sum_{m=1}^M \int_{S_m} \Im \zeta_m |u|^2 ds-k\limsup_{R\to 
\infty}\int_{|x|=R}|u|^2 ds.
\end{equation}
Since all the terms on the right side of this relation are non-positive
by the assumptions \eq{1.14}, it follows that 
$$\limsup_{R\to\infty}\int_{|x|=R}|u|^2 ds=0.$$
This implies that $u=0$ (see,  \cite{R470}, p. 231). 

Thus, uniqueness of the solution to problem \eq{1.5} -- {1.7} is proved.

Let us prove the existence of the solution to \eq{1.5} -- \eq{1.7} of the 
form \eq{1.10}. The existence of the solution of the form \eq{1.10} will be 
established if one proves the existence of $\si_m, 1 \leq m \leq M$, such 
that boundary condition \eq{1.7} is satisfied:
\begin{equation}
	\label{E3.22} u_{eN} - \zeta_j u_e + \frac{A_j \si_j - \si_j}{2}- 
\zeta_j T_j \si_j =0, \quad 1\leq j \leq M.
\end{equation}
Here  $u_e$, which depends on $j$, is defined by the 
formula:
\begin{equation}
	\label{E3.23} u_e:= u - \int_{S_j} G(x,s)\si_j(s)ds=u_0 + 
\sum_{m\neq j}\int_{S_m} G(x,s) \si_m (s)ds.
\end{equation}
Under our assumptions $S_m \in C^{1,\la}$ uniformly with respect to $m$. 
Therefore equation \eq{3.22} is of Fredholm type in the space 
$L^2(\bigcup_{m=1}^M S_m)$. The corresponding homogeneous equation, i.e., 
the equation with $u_0=0$, cannot have a nontrivial solution because such a 
solution would generate by formula \eq{1.10} with $u_0=0$ a function 
$u_M(x)=\sum_{m=1}^M \int_{S_m} G(x,s)\si_m(s)ds$, which would solve the 
homogeneous problem \eq{1.5} -- \eq{1.7}. We have already proved that such a 
function has to be zero in $\R^3 \setminus \bigcup_{m=1}^M D_m$. Thus, 
$u_M\rain{S_m} =0$, $1\leq m \leq M$, and $u_M$ solves the problem:
\begin{equation}
	\label{E3.24}
	L_0 u_M=0 \text{ in } D_m, \quad u_M\rain{S_m} =0, \ 1 \leq m \leq M.
\end{equation}
If $\diam D_m \leq 2a$ is sufficiently small, then problem \eq{3.24} has 
only 
the trivial solution for every $m,1\leq m \leq M$. 
Therefore $u_M=0$ in $D_m$ and in $\R^3\setminus \bigcup_{m=1}^M D_m$. 
Therefore, by the formula for the jumps of the normal derivatives 
of the single layer potential, 
$$\frac{\pa u_M^+}{\pa N}\rain{S_m} - \frac{\pa u_M^-}{\pa N}\rain{S_m} 
=\si_m,$$ 
we conclude that $\si_m=0$, $1\leq m \leq M$. This implies the 
existence of the solution to problem \eq{1.5} -- \eq{1.7} of the form \eq{1.10}.

Theorem \ref{T1} is proved. \qed
$$     $$

Let us return to the assumptions of Theorem 2, namely,
$$\zeta_m=O(\frac 1 a), \quad aN(\Delta_b(y))=N(y) 
|\Delta_b(y)|(1+o(1)),
$$
where $\Delta_b(y)$ is a cube, centered at the point $y$ with the side 
$b>0$, and $o(1)$  is related to the limiting process $b\to 0$.

Under these assumptions let us
establish  an estimate for the function $v_M:=u_M-u_0$, 
which is uniform with respect to $M\to \infty$, or $a\to 0$. From this 
estimate it 
follows that $v_M$ converges, as $a\to 0$, in $L^2(\R^3, (1+|x|)^{-1-\gamma})$, 
where $\gamma>0$ is an arbitrary fixed constant. The function $v_M$
satisfies the radiation condition at infinity. The function 
$u_0\in H^2_{loc}(\R^3)$ solves the equation $L_0 u_0=0$ in $\R^3$.

Let $D_e:=\R^3\setminus \cup_{m=1}^M D_m$ and $S':=\cup_{m=1}^M S_m$.
Let 
$$||v||:=\Big(\int_{D_e}|v(x)|^2(1+|x|)^{-1-\gamma}dx\Big)^{1/2}, \quad 
|||v|||=\sum_{m=1}^M\Big(\int_{S_m}(|v_N|^2+|v|^2)ds\Big)^{1/2}.$$ 

The estimate we wish to prove is:
\begin{equation}
\label{E3.25}
||v_M||\leq c |||u_0|||.
\end{equation}
Here and below $c>0$ stand for various constants independent of $a$.

Let us outline the proof of inequality \eq{3.25}.

{\it Step 1.} If $M=O(\frac 1 a)$, then the right side of \eq{3.25} is 
bounded 
as $a\to 0$.

Indeed, the number of small particles is $M=O(\frac 1 a)$
and $u_0$ is $H^2_{loc}(\R^3)$, so that $u_0$ and $u_{0N}$
are bounded in $L^2(S_m)$ uniformly with respect to $m$,
$1\leq m \leq M$. Thus, 
$$|||u_0|||\leq \frac c {a} \max_{1\leq m \leq 
M}\Big(\int_{S_m}(|u_{0N}|^2+|u_0|^2)ds\Big)^{1/2}\leq \frac c 
{a}|S_m|^{1/2}\leq 
c,$$
where $c>0$ stand for various constants independent of $a$.

{\it Step 2.}  If the inequality \eq{3.25} is false, then there is a 
sequence
$u_{0}^{(n)}$, $|||u_{0}^{(n)}|||=1,$ such that 
$||v_M^{(n)}||:=||v^{(n)}||\geq n$.
 
Define $w^{(n)}:=\frac {v^{(n)}}{||v^{(n)}||}$. Then 
\begin{equation}
\label{E3.26}
||w^{(n)}||=1.
\end{equation}
From the weak compactness of bounded sets in $L^2$, it follows, 
that one may select a subsequence, denoted again
$w^{(n)}$, such that $w^{(n)}$ converges weakly in $L^2_{loc}(D')$
to a function $w$. The function $w^{(n)}$ solves the problem: 
\begin{eqnarray}
\label{E3.27}
\nn L_0 w^{(n)}=0 \quad in \quad D_e, \\ 
w^{(n)}_N-\zeta_m w^{(n)}=(\zeta_m u^{(n)}_{0}-u^{(n)}_{0N})/||v^{(n)}|| 
\quad on \quad S_m, 1\leq m \leq M,
\end{eqnarray}
and $w^{(n)}$ satisfies the radiation condition. 

It follows from \eq{3.27} that $||\na^2w^{(n)}||<c$, so
$||w^{(n)}||_{H^2_{loc}(D_e)}<c$, where $H^2_{loc}(D_e)$ is the Sobolev 
space. Thus, one may assume, using the compactness of the embedding from
$H^2_{loc}$ into $L^2_{loc}$, that $w^{(n)}$ converges to $w$
strongly in $L^2_{loc}(D_e)$. This and equation \eq{3.27} imply
that $w^{(n)}$ converges to $w$ strongly in $H^2_{loc}(D_e)$, so
that $w$ solves equation \eq{3.27}, satisfies the radiation condition and
the homogeneous boundary condition \eq{3.27}, that is,
$w_N-\zeta_m w=0$ on $S_m$.
Therefore, by already proved 
uniqueness theorem (see the proof of Theorem 1), we conclude that $w=0$.
The terms $u^{(n)}_0/||v^{(n)}||$ and $u^{(n)}_{0N}/||v^{(n)}||$ 
tend to zero as $n\to \infty$,
because $||v^{(n)}||>n$. Therefore, the limiting function $w$ satisfies 
the homogeneous boundary condition $w_N=\zeta_m w$ on $S_m$, $1\leq m 
\leq M$.

Let us prove that $|w^{(n)}(x)|<\frac c {|x|},\,\, |x|>R,$ where
$R>0$ is sufficiently large and $c>0$ does not depend on $n$.

For $w^{(n)}$ one has a representation by the 
Green formula in the region
$|x|>R$, where $R>0$ is large enough, so that the ball $B_R:=\{x:\, 
|x|<R\}$ contains $D$. Namely
\begin{equation}
\label{E3.28}
w^{(n)}(x)=\int_{|s|=R}(w^{(n)}_rg(x,s)-g_r(x,s)w^{(n)})ds,\quad |x|>R,
\end{equation}
where the derivatives with respect to $r$ are the derivatives along the 
normal to the sphere $S_R:=\{s: |s|=R\}$, and
$g$ is defined in \eq{1.15}.
It follows from \eq{3.28} that $|w^{(n)}(x)|<\frac c{|x|}$ for 
$|x|>R$, where $c>0$ is a constant independent of $n$, because
local converegence in $H^2$ implies that the $L^2(S_R)$-norms of 
$w^{(n)}$ and of $w^{(n)}_r$ are bounded uniformly with respect to $n$. 

Therefore 
\begin{equation}
\label{E3.29}
\lim_{n\to \infty}||w^{(n)}-w||=0,
\end{equation}
because on compact sets $\lim_{n\to 
\infty}||w^{(n)}-w||_{H^2_{loc}(D_e)}=0$, and near infinity the inequality
$|w^{(n)}(x)|<\frac c{|x|}$ implies that 
$$\int_{\{x: 
|x|>R\}}|w^{(n)}(x)|^2(1+|x|)^{-1-\gamma}=O(R^{-\gamma})\to 0,\,\, R\to 
\infty,$$ 
so that \eq{3.29} holds. Because of the uniqueness of the
limit, not only a subsequence of $w^{(n)}$ but the sequence itself
converges to $w$ as $n\to \infty$. 

This leads to a contradiction, because $w=0$ and 
\eq{3.26} together with \eq{3.29} imply $||w||=1$.

This contradiction proves inequality \eq{3.25}.

From inequality \eq{3.25} and Step 1 one concludes that that
$u_M$ contains a weakly convergent in $L^2_{loc}(D_e)$ subsequence.
By the arguments, similar to the given above, this subsequence
converges in $L^2(\R^3, (1+|x|)^{-1-\gamma})$.
Its limit solves equation \eq{1.23}. 

The relation  $M=O(\frac 1 a)$
plays an important role in our proof of Theorem 2 and in Step 1 in 
the above argument.

\section{Application to creating smart materials}\label{S4}
Let us ask the following question: can one make a material with a
desired refraction coefficient $n(x)$ in a bounded domain $D\subset
\R^3$, filled by a material with a known refraction coefficient
$n_0(x)$, for example $n_0(x) = n_0 = {\rm const}$ in $D$, by embedding
into $D$ a number of small particles, each of which is defined by its
shape and its boundary impedance? 

Consider first the particles
satisfying the assumptions of Theorem \ref{T2}. More specifically,
suppose that all the particles are balls of the same radius $a$. In this case
$$\n S_m\n = 4\pi a^2,\quad \n J_m\n = \int_{\n s\n = a} \int_{\n t\n = 
a} 
\frac{ds\, dt}{\n s-t\n} = 16\pi^2 a^3,$$ so 
$$c_1 = 4\pi, \quad c_2 = 16 \pi^2, \quad \frac{4\pi c_1^2}{c_2} = 
4\pi,$$ 
and formula \eq{1.24} yields 
\begin{equation}
\label{E4.1}
p(x) = \frac{4\pi N(x)\, h(x)}{1 + h(x)}\,,
\end{equation}
where $h(x)$ is defined by the choice of the boundary impedances by
formula \eq{1.20}:
\begin{equation}
\label{E4.2}
\zeta(x) = \frac{h(x)}{a}\,,
\end{equation}
and $N(x)$ is defined by formula \eq{1.21}.
\par
If the original refraction coefficient is $n_0(x),$ then the
corresponding potential is $q_0(x) = k^2[1-n_0(x)]$ by formula \eq{1.2}. 
If the
desired refraction coefficient in $D$ is $n(x)$, then 
the corresponding potential is $q(x) =
k^2[1-n(x)]$, so
\begin{equation}
\label{E4.3}
p(x) = q(x)-q_0(x) = k^2[n_0(x)-n(x)].
\end{equation}
To create a material with the desired refraction coefficient $n(x)$ it
is sufficient to choose $N(x)$ and $h(x)$ so that \eq{4.1} holds with
$p(x)$ defined in \eq{4.3}. If the new material with the refraction
coefficient $n(x)$ has some absorption, that is, ${\rm Im}\, n(x)\geq
0$, and ${\rm Im}\, n_0 = 0$, then ${\rm Im}\, p(x)\leq 0$. Let us prove
that any function $p(x)$ in $D$ with $\Im p\leq 0$, can be obtained 
(in many ways, non-uniquely) by
formula \eq{4.1} with some choices of a nonnegative function $N(x)$ and a
function $h(x)$ with $\Im h \leq 0$. 

Let $p(x) = p_1(x) + ip_2(x)$,
$p_2(x) \leq 0$, and $h(x) = h_1(x) + i h_2(x)$, $h_2(x)\leq 0$. Assume
that $p(x)$ is given. Then \eq{4.1} implies
\begin{equation}
\label{E4.4}
p_1 + ip_2 = 4\pi\, \frac{(h_1 + ih_2)(1 + h_1-ih_2)}{(1 + h_1)^2 +
h_2^2}\, N(x).
\end{equation}
Thus
\begin{equation}
\label{E4.5}
p_1 = 4\pi \, \frac{h_1 + h_1^2 + h_2^2}{(1 + h_1)^2 + h_2^2}\,
N(x),\quad p_2 = 4\pi \, \frac{h_2}{(1 + h_1)^2 + h_2^2}\, N(x).
\end{equation}
There are many choices of the three functions: $N(x) \geq 0$, $h_2(x)
\leq 0$ and a real-valued function $h_1(x)$ such that relations \eq{4.5} 
hold. For example,
if $p_1 > 0$ and $p_2 \neq 0$, then one can choose 
\begin{equation}
\label{E4.6}
h_1(x)= 0,\quad h_2(x) = \frac{p_1(x)}{p_2(x)}\,,\quad N(x) 
=\frac{p_1^2(x) 
+p_2^2(x)}{4\pi \, p_1(x)}\,. 
\end{equation}
It is a simple matter to check that relations \eq{4.5} hold with the
choice \eq{4.6}. Since one has three functions $h_1(x)$, $h_2(x)\leq 0$
and $N(x)\geq 0$ to satisfy two equations \eq{4.5} with $p_2(x)\leq 0$,
there are many ways to do this. A particular choice of $h(x) = h_1(x) +
ih_2(x)$ and $N(x) \geq 0$ yields the surface impedance $\zeta(x)$ of the 
particles
to be embedded around each point $x\in D$, $\zeta(x) = \frac{h(x)}{a}$ by
formula \eq{4.2}, and the number of particles per unit volume around
the point $x$, namely, by formula \eq{1.21} this number is 
$\frac{N(x)}{a}$, so
that the number of particles to be embedded in the volume $dx$ around
point $x$ is equal to $\frac{N(x)}{a}\, dx$. The smallest distance $d$ 
between the
embedded particles should satisfy the inequality $d\gg a$. One may try
to take practically $d > 10a$.
\begin{exa}\label{E1}
{\rm Suppose that the elementary subdomain $\Da_p$, used in the proof of
Lemma \ref{L3}, is a cube with the side $b\gg d$, $x\in \Da_p$. Let, for
example $b = 10^{-2} {\rm cm}$, $d = 10^{-3} {\rm cm}$, $a = 10^{-5} {\rm
cm}$. Then there are $\f \frac{b}{d}\g^3 = 10^{3}$ small particles in
$\Da_p$ around a point $x$, the center of $\Da_p$. 
The function $N(x)$
in $\Da_p$ in this example is found from the formula $\frac{N(x)}{a}\,
b^3 = 10^3$ (use \eq{1.21} with $\tilde{D}=\Delta_p$), so $N(x) = 
10^{-5}\cdot 10^{3} \cdot 10^6 = 10^4$. The
number of small particles, embedded in the cube $\Da_p$ around point
$x$, the center of this cube, is $10^3$ in this example. The relative
volume of these particles in $\Da_p$ is $10^3\cdot \frac{4}{3}\, \pi
10^{-15}\cdot 10^6 = 4.18\cdot 10^{-6}$, so it is quite small, which is
in full agreement with our theory.
\par
The assumption \eq{1.8}, specifically, $d\gg a$, is compatible with the
requirement \eq{1.21}. Indeed, if one denotes by $N(\s{D})$ the left 
side of \eq{1.21}, then  $N(\s{D}) = O\f \frac{1}{a}\g$ for any $\s{D}\seq
D$. 

Let us assume that $\s{D}$ is a unit cube,
and denote by $N(\s{D})$ the left side of \eq{1.21}.
The assumption $d\gg a$ implies that the number $N(\s{D})$ of
particles in $\sD$ is $O\f \frac{1}{d^3}\g$. These relations are
compatible if and only if $O\f \frac{1}{a}\g = O\f \frac{1}{d^3}\g$,
i.e., $d = O(a^{1/3})$. Therefore, it is possible to have $a\ra 0$, 
$\frac{a}{d}\ra 0$ and equation \eq{1.21} satisfied.
\par
Let us discuss the new material properties, specifically, anisotropy, when 
acoustically hard
particles are embedded in the domain $D$, and the assumptions of Theorem
\ref{T3} are valid. The physical situation is now quite different from
the one in Theorem \ref{T2}. From the physical point of view one can 
anticipate the drastic
difference because the wave scattering by one small acoustically soft
particle of the characteristic size $a$ is isotropic and the scattering
amplitude is of order $a$, while the wave scattering by a small
acoustically hard particle is anisotropic and the corresponding
scattering amplitude is of order $k^2 a^3$, (see \cite{R476}, chapter
7).
We assume that $ka\ll 1$, say $ka < 0.1$, so that the quantity
$k^2a^3=(ka)^2a$ is $100$ times less than $a$.
}\end{exa}
\begin{exa}\label{E2}
{\rm Let us assume again that the small particles are all balls of the
same radius $a$. Then 
$$V_m = \frac{4}{3}\, \pi a^3,\quad  \nu(y)\n \Da_p\n=
\frac{4}{3}\, \pi a^3\, N(\Da_p),$$ where $N(\Da_p)$ is the number of
small particles in a small cube $\Da_p$ centered at the point $y$. If
$b$ is the size of the edge of the cube $\Da_p$, then $\nu(y) = 4.18\,
\frac{a^3}{b^3}\, N(\Da_p)$, where $4.18$ is an approximate value of
$\frac{4\pi}{3}\,$. The magnetic polarizability tensor $\be_{pj}$ of a
ball of  radius $a$ is $\be_{pj} = - \frac{3}{2}\, \da_{pj}$, while the
electric polarizability tensor of a perfectly conducting ball is
$3\da_{ij}$, where $$\da_{pj} = \begin{cases} 1, & p=j,\\ 0, & p\neq j.
\end{cases}$$
These values differ by the factor $4\pi$ from the values in \cite{L}
because we use the formula $\vp = \frac{1}{4\pi \n x\n}$ for the
potential of a point charge, while in \cite{L} this potential is
$\frac{1}{\n x\n}\,$. In our example $\be_{pj}$ does not depend on $m$.
Therefore the limit \eq{1.26b} exists if the limit \eq{1.27} exists.
The limit \eq{1.27} exists if and only if the following limit exists:
\begin{equation}
\label{E4.6a}
\frac{4\pi}{3}\, \lim_{a\ra 0}\, a^3 \sum_{D_m\subset \sD} 1 = 
\int_\sD \nu(y)dy,
\end{equation}
where $\nu(y)$ is the function defined in \eq{1.27}. Thus, in contrast
to Example \ref{E1}, where $N(\Da_p) = O\f\frac{1}{a}\g$, now we  have
$N(\Da_p) = O\f \frac{1}{a^3}\g$. The relative volume of the small
particles in Example \ref{E2} is not negligible and does not go to
zero as $a\ra 0$, in contrast to Example \ref{E1}.
}
\end{exa}
\par
Let us discuss the compatibility of the condition
$d\gg a$ and the existence of the limits \eq{1.26b} and \eq{1.27}. If
the condition $d\gg a$ is compatible with the existence of the limit
\eq{1.27}, then it is compatible with the existence of the limit
\eq{1.26b}. If the limit \eq{1.27} exists, then $a^3 N(\s{D}) = O(1)$, so
$N(\s{D}) = O(a^{-3})$. On the other hand, $N(\sD) = O(d^{-3})$. These
relations, in general, are not compatible because $d\gg a$. Let us argue 
more precisely. Let
$\sD = \Da_p$, where $\Da_p$ is a cube with the edge of size $b$. Let us
assume that the small particles in $\Da_p$ are identical and their
characteristic size is $a$. If \eq{1.27} holds, where $\nu(y)$ is
continuous, and if $b$ is small, then the right side of \eq{1.27} equals
to $\nu(y) b^3$, $y\in \Da_p = \sD$. The left side of \eq{1.27} equals
to $c_3 a^3 N(\Da_p)$. Thus $N(\Da_p) = \frac{1}{c_3}\, \nu(y)\,
\frac{b^3}{a^3}\,$. On the other hand, $N(\Da_p) = \frac{b^3}{d^3}\,$,
prvided that one assumes that the centers of the small particles are at 
the uniform grid, so that there are $\frac{b}{d}$ centers on the segment 
of length $b$. If $\frac{1}{c_3}\, \nu(y)\, \frac{b^3}{a^3} =
\frac{b^3}{d^3}\,$, then $\frac{a}{d} = \f \frac{\nu(y)}{c_3}\g^{1/3}$.
Therefore the condition $d\gg a$ is satisfied only if $\f
\frac{\nu(y)}{c_3}\g^{1/3}\ll 1$, say $\f \frac{\nu(y)}{c_3}\g^{1/3}
\leq 0.1$. The number $c_3$ depends on the shape of the particle. If the
particles are balls of radius $a$, then $c_3 = 4.18$. Therefore $\nu(y)
\leq 4.10^{-3}$. 

The conclusion is: 

{\em The condition $d\gg a$ is
compatible with the existence of the limit \eq{1.27} only if the
function $\nu(y)$ in \eq{1.27} is sufficiently small.} 

In general,
equation \eq{1.29} cannot be reduced to a local differential equation
for $\U(x)$. However, if $\nu(y)$ is small, one may use perturbation
theory to study equation \eq{1.29}. However, under an additional
assumption, reasonable from the physical point of view, one can reduce
integral-differential equation \eq{1.29} to a differential equation.
Namely, let us assume that $\nu(y)$ is a continuously differentiable
function in $D$ which vanishes near the boundary $S$.
\par
Under this assumption one can integrate by parts the last integral in 
\eq{1.29} and get:
\begin{equation}
\label{E4.7}
\U(x) = u_0(x) + \int_D G(x,y) \Bka \Da \U(y)\nu(y) + \sum^3_{p,j=1}
\frac{\pa}{\pa y_p}\Ba \frac{\pa \U(y)}{\pa y_j}\, \be_{pj}(y)\nu(y)\Bz\Bkz.
\end{equation}
Let us apply the operator $L_0 = \na^2 + k^2 - q_0(x)$ to \eq{4.7} and
use \eq{1.4} to get:
\begin{equation}
\label{E4.8}
[\na^2 + k^2 - q_0(y)] \U + \nu(y) \na^2 \U(x) +
\sum^3_{p,j=1}\frac{\pa}{\pa y_p}\Ba \frac{\pa \U(x)}{\pa y}\,
\be_{pj}(y)\, \nu(y)\Bz=0,
\end{equation}
where $\U(x)$ satisfies the radiation condition of the type \eq{1.6}.
This is an elliptic equation and the perturbation $\PE$ of the operator
$L_0$ is:
\begin{equation}
\label{E4.9}
\PE\U := \nu(x)\na^2\U(x) + \sum^3_{p,j=1} \frac{\pa}{\pa y_p}\Ba
\frac{\pa \U}{\pa y_j}\, \be_{pj}(y)\, \nu(x)\Bz.
\end{equation}
This perturbation is the sum of the terms with positive small coefficient 
$\nu(y)$ in
front of the second derivatives of $\U$ and a term with the first order 
derivatives of $\U$:
\begin{equation}
\label{E4.10}
\PE\U = \nu(x) \Bka \na^2\U(x) + \sum^3_{p,j=1} \frac{\pa}{\pa x_p}\,
\Ba \frac{\pa \U(x)}{\pa x_j}\, \be_{pj}(x)\Bz\Bkz + \sum^3_{p,j=1}
\frac{\pa \U(x)}{\pa x_j}\, \be_{pj}(x)\, \frac{\pa \nu(x)}{\pa x_p}\,.
\end{equation}
If both $\nu(x)$ and $\na \nu(x)$ are small, this equation can be
studied by perturbation methods. The physical effect on the properties
of the new material, created by embedding into $D$ small acoustically hard 
particles,
consists in appearing of anisotropy in the new material: the propagation
of waves is described by the integral-differential equation \eq{4.7} or
(under the additional assumption on $\nu(y)$, namely: $\nu(y)$ vanishes
near the boundary $S$ of $D$) by the differential equation \eq{4.8} with
variable coefficients in front of the senior (second order) derivatives
and the terms with the first order derivatives.
\par
The role of the compatibility of the assumption $d\gg a$ and of the
assumption \eq{1.27} is quite important. Although passing to the limit
$a\ra 0$, justified in the proof of Theorem \ref{T3}, is based on the
assumptions \eq{1.26b} and \eq{1.27}, but without the assumption $d\gg a$
one cannot expect, in general, that the effective field $u_e(x)$, acting
on any single particle, is practically constant on the distances of the 
order $2a$. This physical assumption is important for our theory.

From the mathematical point of view, 
if $\nu(x)$ is not sufficiently small, then the existence of the unique
solution to equation \eq{4.7} or of the solution to equation \eq{4.8},
satisfying the radiation condition, is not guaranteed.

If, on the other hand, the quantity  
$$\sup_{x\in \R^3} \big(|\nu(x)| + |\na \nu(x)| \big) \ll 1,
$$ 
that is, this quantity is sufficiently small, then one can argue that
the norm of the integral operator in \eq{4.7} in $L^2(D)$ is small, so
that equation \eq{4.7} has a unique solution in $L^2(D)$. This
solution admits a natural extension to the whole space $\R^3$ by the
right side of \eq{4.7} because $\nu(y)$ vanishes outside $D$. Since
$G(x,y)$ satisfies the radiation condition, the solution to \eq{4.7}
also satisfies this condition. Without the assumption that $\n \nu(x)\n
+ \n \na \nu(x)\n$ is sufficiently small, one cannot use the above
argument. 

With this assumption one may solve equation \eq{4.7} by 
iterations and find in this way an
approximate solution to this equation.
The first iteration yields the following approximate solution to
equation \eq{4.7}:
\begin{equation}
\label{E4.11}
\U(x) = u_0(x) + \int_D G(x,y)\, \Bka \Da u_0(y) \nu(y) + \sum^3_{p,j=1}
\frac{\pa}{\pa y_p} \Ba \frac{\pa u_0(y)}{\pa y_j}\, \be_{pj}(y)\,
\nu(y)\Bz \Big] dy.
\end{equation}
Formula \eq{4.11} gives the correction to the solution $u_0(x)$ of the
unperturbed scattering problem, i.e., the scattering problem in the
absence of small bodies. Since one has 
$$\Da u_0 = -k^2 n_0(x)u_0,$$ 
\eq{4.11} can be rewritten as:
\begin{align}
\nn \U(x) =   u_0(x) & - k^2 \int_D G(x,y) n_0(y) u_0(y)\nu(y) dy  \\
\label{E4.12}  \mbox{} & +  \int_D G(x,y) \sum^3_{p,j=1} \frac{\pa}{\pa
y_p} \Ba \frac{\pa u_0(y)}{\pa y_j}\, \be_{pj}(y)\nu(y)\Bz dy.
\end{align}
In \cite{MK}, Chapter 3, Section 3, the Neumann problem for the Helmholtz 
equation with 
$n_0(x)=1$ was studied in the domain, similar to the one in  
equation \eq{1.5} and it was proved under the assumptions used in 
\cite{MK}, that the main term of the asymptotics
of the solution, as the relative volume of the particles tends to zero,  
is the incident 
field, while the next term is proportional to this relative volume.


\end{document}